\documentclass[aps,prl,reprint,superscriptaddress,floatfix,nofootinbib,showkeys,showpacs,preprintnumbers]{revtex4-2}
\pdfoutput=1

\usepackage{tabularx}
\usepackage{xspace}
\usepackage{listings}
\usepackage{lineno}
\usepackage{bm}
\usepackage{amsmath}
\usepackage{amssymb}
\usepackage[dvipsnames]{xcolor} 
\usepackage{graphicx}
\usepackage{comment}
\usepackage{natbib}
\usepackage{hyperref}
\usepackage{amsmath}
\usepackage{enumitem}
\setlist{wide, labelwidth=!, labelindent=0pt}
\usepackage{multirow}
\hypersetup{    
  colorlinks      = true,
  linkcolor       = {black},
  linkbordercolor = {white},
  citecolor       = {black},
  citebordercolor = {white},
  urlcolor        = {blue},
  urlbordercolor  = {white},
}

\usepackage{soul} 
\usepackage{amsmath}
\usepackage{amssymb}
\usepackage{xspace}
\usepackage{xifthen}
\usepackage{float}
\usepackage{mathtools}
\usepackage{relsize}
\usepackage{aas_macros}

\newcommand{\redmapper}{\textsc{redMaPPer}}

\makeatletter 
\newcommand{\LCASES}[1]{$\m@th\displaystyle{#1}$\hfil}
\newcommand{\CCASES}[1]{\hfil$\m@th\displaystyle{#1}$\hfil}
\newcommand{\RCASES}[1]{\hfil$\m@th\displaystyle{#1}$}
\makeatother

\newcases{ecases}{\quad}{\CCASES{##}}{\LCASES{##}}{\lbrace}{.}
\newcases{ecases*}{\quad}{\CCASES{##}}{{##}\hfil}{\lbrace}{.}

\def\beq{\begin{eqnarray}}
\def\eeq{\end{eqnarray}}

\def\siglogm{\sigma_{\log M}}
\def\Mmin{M_\mathrm{min}}
\def\Msun{M_\odot}

\def\redmapper{redMaPPer}
\def\Rlambda{R_\lambda}

\def\pimax{\Pi_\mathrm{max}}
\def\Ncyl{N_\mathrm{cyl}}

\defcitealias{DESY1CL_2020_et_al}{DES-Y1 clusters analysis} 

\newcommand{\avg}[1]{\langle #1 \rangle}

\begin{document}

\preprint{APS/123-QED}

\title{Dark Energy Survey Year 1 Clusters are Consistent with {\it{Planck}}}

\author{Andr\'{e}s N. Salcedo}
\email[]{ansalcedo@arizona.edu}
\affiliation{ Department of Astronomy/Steward Observatory, University of Arizona, 933 North Cherry Avenue, Tucson, AZ 85721, USA.}
\affiliation{Department of Physics, University of Arizona, 1118 East Fourth Street, Tucson, AZ 85721, USA.}
\author{Hao-Yi Wu}
\affiliation{Department of Physics, Boise State University, Boise, ID 83725, USA}
\author{Eduardo Rozo}
\affiliation{Department of Physics, University of Arizona, 1118 East Fourth Street, Tucson, AZ 85721, USA.}
\author{David H. Weinberg}
\affiliation{Center for Cosmology and AstroParticle Physics (CCAPP), the Ohio State University, Columbus OH 43210, USA}
\affiliation{Department of Astronomy, the Ohio State University, Columbus, OH 43210, USA}
\author{Chun-Hao To}
\affiliation{Center for Cosmology and AstroParticle Physics (CCAPP), the Ohio State University, Columbus OH 43210, USA}
\affiliation{Department of Physics, the Ohio State University, Columbus, OH 43210, USA}
\affiliation{Department of Astronomy, the Ohio State University, Columbus, OH 43210, USA}
\author{Tomomi Sunayama}
\affiliation{ Department of Astronomy/Steward Observatory, University of Arizona, 933 North Cherry Avenue, Tucson, AZ 85721, USA.}
\author{Andy Lee}
\affiliation{Department of Physics, Boise State University, Boise, ID 83725, USA}
\affiliation{Department of Physics, the Ohio State University, Columbus, OH 43210, USA}

\date{\today}

\begin{abstract}
 
The recent Dark Energy Survey Year 1 (DES-Y1) analysis of galaxy cluster abundances and weak lensing produced $\Omega_{\rm m}$ and $\sigma_8$ constraints in 5.6$\sigma$ tension with {\it{Planck}}.  It is suggested in that work that this tension is driven by unmodelled systematics in optical cluster selection.  We present a novel simulation-based forward modeling framework that explicitly incorporates cluster selection into its model predictions. Applying this framework to the DES-Y1 data we find consistency with {\it{Planck}}, resolving the tension found in the DES-Y1 analysis.  An extension of this approach to the final DES data set will produce robust constraints on $\Lambda$CDM parameters and correspondingly strong tests of cosmological models.

\end{abstract}

\maketitle

\noindent \textbf{\emph{Introduction.}} --- The Dark Energy Survey (DES) is a photometric survey designed to place cosmological constraints on dark energy using a variety of cosmological probes \cite{DES_2005}.  Perhaps the most surprising result from the DES thus far is the strong tension between the cosmological parameters inferred from the DES Year 1 (DES-Y1) galaxy clusters \cite{DESY1CL_2020_et_al} and those inferred from both {\it Planck} \cite[$5.6\sigma$,][]{Planck_2016} and the DES 3$\times$2pt analysis \cite[$2.4\sigma$,][]{DES_3x2pt_2018,DES_3x2pt_2021}. The DES-Y1 cluster paper argued these tensions were likely driven by errors in the modeling of the weak lensing signal of the \redmapper\ clusters, a hypothesis that has been confirmed by a variety of works \citep{Costanzi_et_al_2020,To_et_al_2021a}.

We develop a novel method for forward-modeling cluster selection in simulations, and apply it to the DES-Y1 cluster data set. Our method builds on the work of \cite{Wu_et_al_2022}, who demonstrated that the impact of projection effects on cluster selection can be modeled using a counts-in-cylinders approach.  We apply this scheme to simulation-based mock galaxy catalogs created using a Halo Occupation Distribution (HOD) framework, fitting for the HOD parameters at fixed {\it Planck} cosmology. Our model fits the DES-Y1 cluster abundance and weak lensing data when applied to the {\it Planck} cosmology simulations from the \textsc{Abacus Summit} simulation suite \cite{Maksimova_Summit_et_al_2021}, thereby resolving the tension found in the DES-Y1 analysis. 

\noindent \textbf{\emph{Data and measurement.}} --- We use measurements of the lensing and abundance of 6504 redMaPPer \cite{Rykoff_et_al_2014} clusters in the $1321 \, \mathrm{deg}^2$ of DES-Y1 imaging data \cite{Drlica-Wagner_et_al_DESY1_2018}.  Our cluster sample is separated into three redshift bins, $0.20 < z < 0.35$, $0.35 < z < 0.50$, and $0.50 < z < 0.65$. Each redshift bin is further separated into four bins of richness $\lambda$. For each of these bins, cluster shear profiles were measured using the DES-Y1 {\sc{Metacalibration}} shape catalog \cite{Zuntz_et_al_DESY1_Metacal_2018} and the BPZ photometric redshift catalog \cite{Hoyle_et_al_DESY1_BPZ_2018}, then boost factor corrected using the results of \cite{Varga_et_al_DESY1_2019}. The full details of this measurement are found in \cite{McClintock_DES_2019}. \\ 

\noindent \textbf{\emph{Cluster modeling.}} --- We randomly assign galaxies to halos in the simulation using an HOD framework.  Clusters are identified as cylindrical overdensities in the galaxy counts. The details of this procedure are specified below. \\

{\emph{Simulations and halo identification }}
---
We use the halo and particle data from the AbacusSummit \cite{Maksimova_Summit_et_al_2021} suite of N-body simulations.\footnote{https://abacussummit.readthedocs.io/en/latest/index.html} We use 6 realizations of their fiducial {\it Planck} cosmology \citep{Planck_2016}, corresponding to a flat $\Lambda$CDM model with $\Omega_m = 0.314$, $h=0.6736$, $\sigma_8 = 0.8080$, and $n_s = 0.9649$.  All simulations used periodic cubes with side-length $L_\mathrm{side} = 2.0 \, h^{-1} \, \mathrm{Gpc}$, $N_\mathrm{part} = 6192^3$ particles of mass $M_\mathrm{part} \approx 2 \times 10^9 \, h^{-1} \, M_\odot$, and spline force softening length $\epsilon_g = 7.2 \, h^{-1} \, \mathrm{kpc}$ \cite{Garrison_et_al_2018}. Halos are identified from particle snapshots using the {\sc{CompaSO}} halo finder \cite{Hadzhiyska_COMPASO_et_al_2022}. We use the ``cleaned'' {\sc{CompaSO}} halo catalogs and adopt as the halo center the center-of-mass of the most massive subhalo as recommended in \cite{Hadzhiyska_COMPASO_et_al_2022}. In what follows, we will denote {\sc{CompaSO}} halo masses as $M_{\rm h}$. We use simulation redshift snapshots at $z = 0.3$, $0.4$, and $0.5$ to model the lensing in the three DES-Y1 redshift bins. To account for the small difference between snapshot outputs and the median cluster redshift in each bin, we map all halo masses from the snapshot redshift onto the clusters' median redshift via abundance matching using the \cite{Tinker_et_al_2008} mass function.

{\emph{Halo occupation modeling}} --- Halos in the simulation are populated with galaxies using an HOD framework \cite[e.g.][]{Berlind_2002}.  Following standard practice \cite{Zheng_et_al_2005} we separate galaxies into satellites and centrals and parameterize their respective mean occupations as
\begin{align}
\langle N_\mathrm{cen} | M \rangle &= \frac{1}{2} \left[ 1 + \mathrm{erf} \left( \frac{\log_{10} M - \log_{10} \Mmin}{\siglogm} \right) \right],\\
\langle N_\mathrm{sat} | M \rangle &= 
\begin{ecases*}
     \left( \frac{M - M_0}{M_1} \right)^\alpha& if $N_\mathrm{cen}=1$, \\
     0 & otherwise.
\end{ecases*}
\end{align}

In practice, we present results in terms of the parameter $N_{14}$, the mean number of galaxies in a $10^{14}\ h^{-1}\ \Msun$ halo.  We favor $N_{14}$ over $M_1$ as the former is more appropriate for the characteristic mass scale of the \redmapper\ systems. 

The actual number of centrals placed in a halo is a random draw from the Bernoulli distribution with mean $\avg{N_{\rm cen}|M}$. Each halo with a central galaxy is also assigned a number of satellite galaxies drawn from a Poisson distribution with mean $\langle N_\mathrm{sat} | M \rangle$. We have also considered models with an additional log-normal scatter, but have found that the resulting scatter parameter is always consistent with zero. Centrals are placed at the center of their host halo, while satellites are distributed according to an NFW profile \cite[NFW;][]{NFW_1997} using the concentration--mass relation of \cite{Correa_2015}. Because our data vector is insensitive to $M_0$ and $\siglogm$, we fix these parameters to $\log M_0 = 11.7$ and $\siglogm = 0.1$.\\

{\emph{Counts-in-cylinders as an optical richness proxy}} --- At a given cluster mass the scatter in {\redmapper} richness and lensing are positively correlated. Consequently, the lensing of richness-selected clusters is higher than that predicted from their halo masses alone.  This selection bias can be attributed to projection effects: halos are more likely to be included in a cluster catalog if their respective lines-of-sight are over-dense, which boosts both the cluster richness and weak lensing signal. Several works have shown that these effects can be qualitatively reproduced by modeling cluster richness selection with a simple counts-in-cylinders process \cite{Angulo_et_al_2012, Busch_White_2017, Sunayama_et_al_2020, Sunayama_More_2019, Wu_et_al_2022, Sunayama_et_al_2023, Zeng_et_al_2023}.

We extend the counts-in-cylinders approach of \cite{Wu_et_al_2022} by weighting each HOD galaxy according to its line-of-sight distance from the central halo as per the projection model of \cite{Costanzi_et_al_2019a}:
\beq
    w(d_\mathrm{los}, d_{\rm cyl}) =
    \begin{ecases*}
    1 - \left(\frac{d_\mathrm{los}}{d_{\rm cyl}}\right)^2 & if $|d_\mathrm{los}| < d_{\rm cyl}$, \\
    0 & otherwise.
    \end{ecases*}
\eeq
Here, $d_\mathrm{los}$ refers to the line-of-sight separation of a simulated galaxy and the cluster halo center, and the depth $d_\mathrm{cyl}$ is the line-of-sight separation beyond which galaxies no longer contribute to the cluster richness. We define $N_{\rm cyl}$ as the sum of the galaxy weights of all galaxies within the cylinder, and within a projected aperture
\begin{align}
\Rlambda &= \left( \frac{\Ncyl}{100} \right)^{0.2} \, \left[ h^{-1} \,  \mathrm{physical} \, \mathrm{Mpc} \right]
\end{align}
that mimics that of the \redmapper\ algorithm. To generate a cluster catalog, we percolate the galaxy catalog by demanding that each galaxy belong to at most one cluster, with preference given to the most massive system.  We emphasize that $\Ncyl$ is \it not \rm equivalent to the {\redmapper} richness $\lambda$ but rather is a richness proxy.  

The relation between $\lambda$ and $\Ncyl$ is model-dependent, and is determined empirically using abundance matching. Specifically, we demand
\beq
n_{\rm halos}(N_{\rm cyl}) = n_{\rm clusters}(\lambda),
\eeq
where $n_{\rm halos}$ and $n_{\rm clusters}$ are the cumulative number density of halos and clusters in our fiducial cosmology. The end product of our algorithm is a simulation in which: 1) every halo has been assigned a richness $\lambda$ based on its line-of-sight galaxy counts $N_{\rm cyl}$; and 2) the number density of clusters in the simulation exactly reproduces that of the data.
\\

\noindent \textbf{\emph{Emulation and likelihood analysis.}}--- 
Following \cite{McClintock_DES_2019}, we adopt a Gaussian likelihood for the lensing profile of the galaxy clusters. We also adopt the semi-analytic covariance matrix used in that work.  However, we use the lensing profiles measured from our simulations rather than their analytic prescriptions as the theoretical value in our likelihood analysis.
\\

\begin{table*}[]
    \centering
    \caption{Initial sampling, prior ranges, and posterior constraints of model parameters from DES-Y1 data. In all cases the initial sampling range used to train the emulator is used as a uniform prior during MCMC analyses. Values quoted are posterior medians with $68\%$ confidence intervals. The value of $\chi^2 / \mathrm{d.o.f.}$ is calculated from the mean of the posterior samples.  The parameters $\epsilon$ all refer to the coefficient characterizing the redshift evolution of the respective HOD parameter (included as a subscript), which is modeled as linear in redshift about the pivot redshift.}
    \begin{tabular}{lccccccccc}
    \hline 
          {\bf{HOD Model Runs}} &  $\log_{10} \Mmin$ & $\log_{10} N_{14}$ & $\alpha$ & $\epsilon_\mathrm{min}$ & $\epsilon_{14}$ & $\epsilon_\alpha$ & $\chi^2/\mathrm{d.o.f}$ & $p$ \\
    \hline 
        Initial Sampling Range & $[11.2,12.8]$ & $[0.0, 1.4]$ & $[0.5, 2.0]$ & - & - & - & - & - \\
    \hline    
    Joint with no evolution & $12.47^{+0.09}_{-0.10}$ & $0.45^{+0.08}_{-0.07}$ & $1.40^{+0.27}_{-0.20}$& - & - & - & $145.66/129$ & $0.150$\\
    Joint with evolution & $12.48^{+0.13}_{-0.16}$ & $0.53^{+0.09}_{-0.09}$ & $1.16^{+0.18}_{-0.15}$ & $-0.32^{+0.83}_{-0.85}$ & $-0.52^{+0.66}_{-0.64}$ & $1.74^{+0.81}_{-1.03}$ & $140.65/126$ & $0.176$ \\
    \hline
    {\bf{Gaussian Model Runs}} & $\sigma$ &  $\log_{10} M_1$ & $\alpha$ & $\epsilon_\sigma$ & $\epsilon_1$ & $\epsilon_\alpha$ & & \\
    \hline
    Initial Sampling Range & $[0.0, 0.8]$ & $[12.5, 14.5]$ & $[0.5, 1.5]$ & - & - & - & - & - \\
    \hline
    Joint with no evolution &  $0.13^{+0.06}_{-0.05}$ & $13.93^{+0.06}_{-0.06}$ & $1.30^{+0.10}_{-0.12}$ & - & - & - & $185.57/129$ & $0.001$ \\
    Joint with evolution & $0.15^{+0.08}_{-0.07}$ & $13.85^{+0.09}_{-0.10}$ & $1.21^{+0.15}_{-0.11}$ &  $-0.07^{+0.41}_{-0.42}$ & $0.47^{+0.64}_{-0.65}$ & $0.60^{1.13}_{-0.85}$ & $179.59/126$ & $0.001$ \\
    \hline
    \end{tabular}
    \label{tab:prior_constraints}
\end{table*}

{\emph{Emulating cluster lensing}}
---
We use {\sc{corrfunc}} \cite{Sinha_2017} to compute the real-space cluster--matter cross-correlation $\xi_{cm}(r_p, \pi)$ for each of our richness bins. The lensing observable $\Delta \Sigma$ is then calculated via
\beq
    \Delta \Sigma (r_p) = \rho_m \left[ \frac{2}{r^2_p} \int_0^{r_p} r' w_{p,cm}(r') d r' - w_{p,cm}(r_p)\right],
\eeq
where $\rho_m = \Omega_m \rho_\mathrm{crit}$ is the cosmological matter density, and $w_{p,cm}$ is the projected cluster--matter cross-correlation function,
\beq
w_{p,cm}(r_p) = 2 \int_0^{\pimax} \xi_{cm}(r_p, \pi),
\eeq
where $\pimax = 100.0 \, h^{-1} \, \mathrm{Mpc}$.

We incorporate miscentering using the model of \citet{Zhang_DES_et_al_2019} by miscentering a fraction $f_\mathrm{offset}$ of our clusters by $r_\mathrm{offset}$, where the latter is drawn from a Gamma distribution,
\beq
P(x|\tau) = \frac{x}{\tau^2} \mathrm{exp}\left( - \frac{x}{\tau} \right),
\eeq
where $x = r_\mathrm{offset} / R_\lambda$. We adopt the best-fitting values from the DES analysis in \cite{Zhang_DES_et_al_2019}, namely $f_\mathrm{offset} = 0.165$ and $\tau = 0.166$. Omitting this correction makes little difference to our results.

We characterize the dependence of $\Delta \Sigma$ on HOD parameters using the Gaussian process emulation scheme of \cite{Wibking_et_al_2020} (also see \cite{Salcedo_et_al_2022}), which uses a squared-exponential kernel in each radial bin of $\Delta \Sigma$. In each bin the hyperparameters of the kernel are obtained by maximizing the leave-one-out cross-validation pseudo-likelihood. For further details we direct the reader to Appendix C in \cite{Wibking_et_al_2020}.

We construct our emulator with a training set generated by a Latin hypercube sampling of the flat priors on our HOD parameters (specified in Table~\ref{tab:prior_constraints}). We compute $\Delta \Sigma$ profiles from our simulations at each point and then train the Gaussian process emulator to predict $\Delta\Sigma$ everywhere else. We then generate an MCMC posterior of the data and randomly select an additional 50 points from the chain to augment our training data. We retrain the emulator, and the full analysis is rerun. The final systematic uncertainty in the emulator profiles is $\lesssim 2\%$, compared to the $10{-}20\%$ errors in the data.

{\emph{Gaussian likelihood model}} --- We use the same likelihood as \cite{McClintock_DES_2019}.  In that work, the authors measured $\Delta\Sigma$ for each richness bin in 11 logarithmically spaced radial bins between $0.2 < r_p < 30.0 \, h^{-1}\ \mathrm{physical-Mpc}$.  The covariance matrix is non-diagonal across radial bins, but different richness bins are assumed to be uncorrelated. This is a reasonable approximation because the errors are shape-noise-dominated. We refer the reader to \cite{McClintock_DES_2019} for further details. The model predictions for $\Delta\Sigma$ are taken directly from the emulator described in the previous section. We use the  {\sc{emcee}} python module \cite{Foreman-Mackey_et_al_2016} to sample our parameter posteriors.\\
\\

  \begin{figure*}
\centering \includegraphics[width=1.0\textwidth]{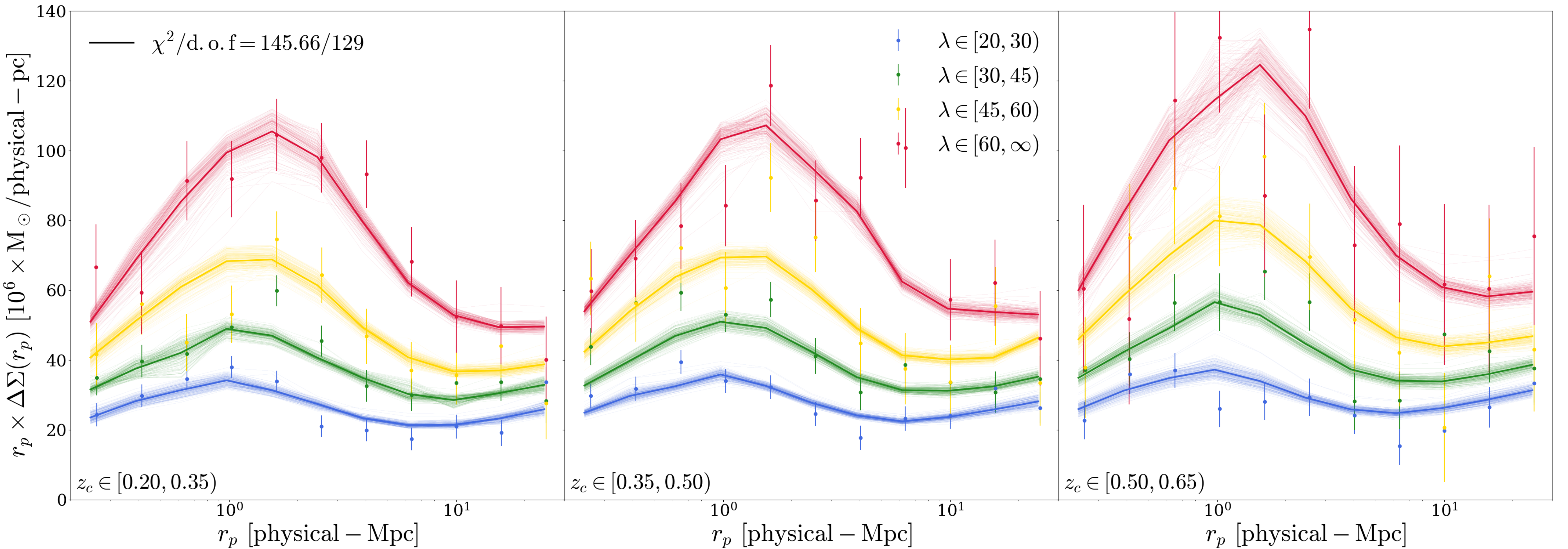}
    \caption{Comparison of DES-Y1 cluster weak lensing profiles (points with errorbars) in redshift bins $z \in [0.20, 0.35)$ (left panel), $z \in [0.30, 0.50)$ (middle panel) and $z \in [0.50, 0.65)$ (right panel) with those predicted by our posterior mean model (thick solid lines) and 200 random samples from our MCMC chain. In all redshift bins our model is run assuming a {\it{Planck}} cosmology with abundances fixed to those measured in the DES-Y1 {\redmapper} sample. We see that in all redshift bins the data is well-described by our model, indicating consistency between DES-Y1 cluster weak lensing, abundances, and {\it{Planck}} cosmology.} 
\label{fig:best-fits}
\end{figure*}

\noindent \textbf{\emph{Results.}} ---  Our fiducial analysis uses a single set of HOD parameters to describe all three redshift bins. Allowing for the HOD parameters to evolve with redshift, we find the evolution parameters are only weakly constrained and are always consistent with zero. Consequently, allowing for redshift evolution has no impact on our conclusions.  For completeness, Table~\ref{tab:prior_constraints} includes the posteriors obtained when allowing for redshift evolution in our model. We further fix $d_\mathrm{cyl}$ to the width of the projection kernel measured in \cite{DESY1CL_2020_et_al}. We have run tests with $d_\mathrm{cyl}$ free and found it to be poorly constrained. Figure~\ref{fig:best-fits} compares the DES weak lensing data to the weak lensing profiles obtained by sampling our model posteriors.  Our model gives an acceptable description of the data, with $\chi^2 / \mathrm{d.o.f} = 145.66/129$ (PTE=15\%). This is in marked contrast with the results of \cite{DESY1CL_2020_et_al}, where the lensing and abundance data could not be simultaneously fit assuming the best-fit DES-Y1 3$\times$2pt cosmology \cite{DES_3x2pt_2018}.  The fact that we can simultaneously fit the abundance and weak lensing signal of DES-Y1 \redmapper\ clusters is the key result in this work.

Our parameters posteriors are summarized in Table \ref{tab:prior_constraints}. Interestingly, we see that: 1) the parameter $\log_{10} \Mmin$ is well constrained; and 2) the posterior value of $\log_{10} N_{14}$ is surprisingly low ($N_{14}\approx 3$) compared to the expected richness of $M = 10^{14} \, h^{-1} \, \mathrm{M_{\odot}}$ clusters ($\lambda\approx 20$).  These results highlight the importance of projections in our model. For the full parameter posterior see \cite{supp}.

\begin{figure*}
\centering \includegraphics[width=1.0\textwidth]{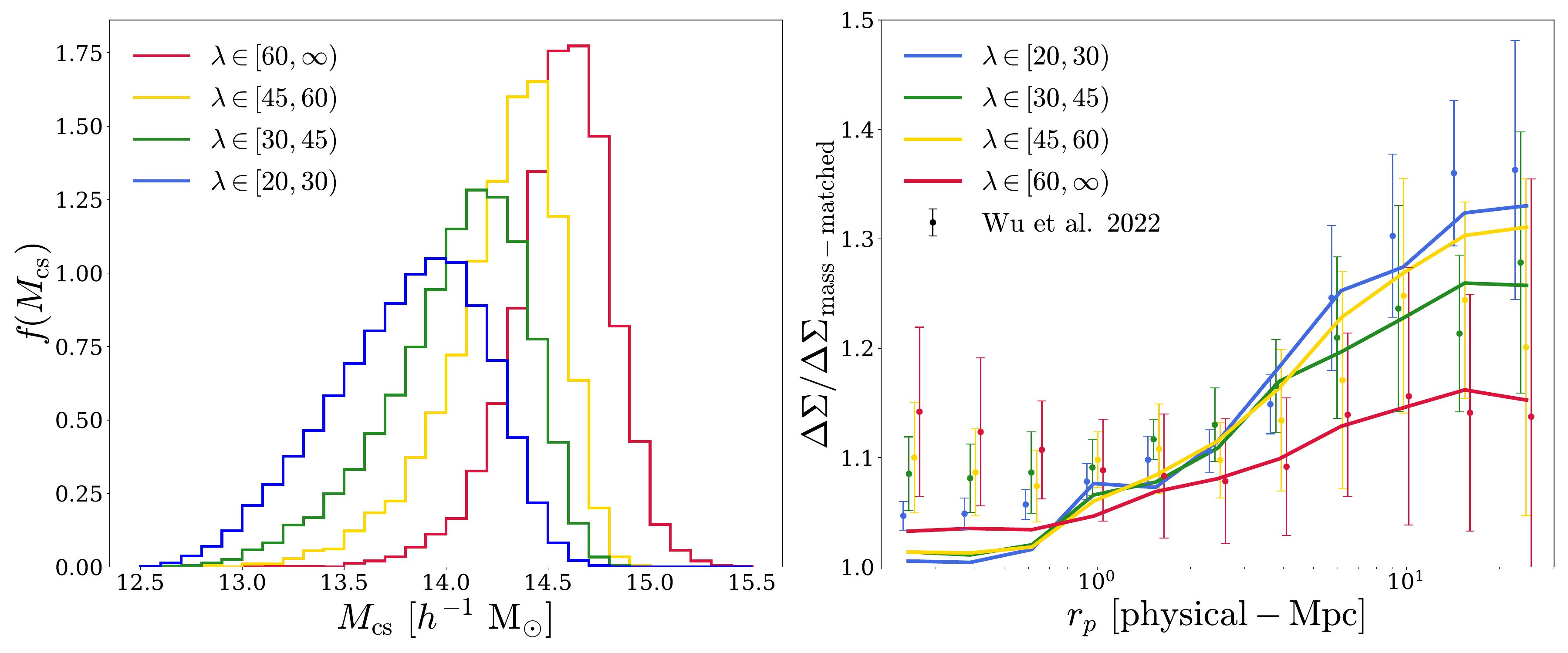}
   \caption{Predictions for our best-fit model to the DES-Y1 cluster weak lensing data at $z \in [0.20, 0.35)$. In the left panel we plot the mass distribution for each richness bin (blue, yellow, green, and red for $\lambda \in [20,30)$, $[30,45)$, $[45,60)$ and $[60, \infty)$ respectively). In particular, we see that our projection model produces long tails to low halo mass in each richness bin. In the right panel we plot the impact of projection effects on our lensing data vector compared to the results from \cite{Wu_et_al_2022} and observe good agreement.} 
\label{fig:mass-rich}
\end{figure*}

Our analysis also allows us to derive several additional quantities of interest. The left panel of Figure~\ref{fig:mass-rich} shows the predicted halo mass distributions for each of our richness bins in the redshift bin $z\in[0.20,0.35]$.  In each richness bin we observe a unimodal mass distribution with a significant low-mass tail. However, we see no evidence of bimodality, as assumed for instance in \cite{Grandis_et_al_2021}.  That is, clusters cannot be unambiguously split into ``clean'' and ``projected'' samples: the impact of projection effects on galaxy clusters forms a continuum in halo mass.

We define the weak lensing bias due to cluster selection as the ratio of our best-fit weak lensing profiles to those obtained from a random sample of halos distributed as per Figure~\ref{fig:mass-rich}.  The lensing bias is small at small scales but grows to 15\%-40\% at large scales.  Moreover, lensing bias increases with decreasing richness.  A comparison of these results to those of \citet{Wu_et_al_2022} is shown in the right panel of Figure~\ref{fig:mass-rich}.

We can also use our posterior distributions to infer the mass--richness relation of DES-Y1 \redmapper\ clusters.  For a comparison to a variety of values from the literature see \cite{supp} (and references therein, \cite{Baxter_et_al_2016, Farahi_et_al_2016, Raghunathan_et_al_2019, Bleem_et_al_2020, DESY1CL_2020_et_al, Costanzi_et_al_2020, To_et_al_2021b}).  Here, we wish to highlight the tension between our results and those of \cite{To_et_al_2021b}.  One possible explanation for the origin of this tension is that we include small scales in our lensing analysis, while \cite{To_et_al_2021b} do not.  However, restricting ourselves to large scales only ($r_p \geq 2.0 \, h^{-1} \, \mathrm{Mpc}$) fails to resolve this tension.  Consequently, either this tension is sourced by the difference in the best-fit cosmologies of the two works, or at least one of the two models fails to properly account for cluster selection effects.

For completeness, we have also performed an analysis similar to that of \cite{DESY1CL_2020_et_al}.  We assumed a power-law richness--mass relation of slope $\alpha$ and normalization $M_1$. The variance in richness is set by the sum in quadrature of the Poisson variance and an intrinsic lognormal scatter parameter $\sigma$.  We randomly assigned richness values to simulated halos and used emulators to characterize the dependence of the resulting lensing signal as a function of $\alpha$, $\log M_1$, and $\sigma$. Our results are summarized in Table \ref{tab:prior_constraints} (Gaussian Model Runs). As expected, this model cannot describe the abundance and lensing signal of DES-Y1 clusters in a {\it{Planck}} cosmology (PTE=$0.1\%$). \\

\noindent \textbf{\emph{Conclusions.}} --- The recent DES-Y1 analysis of the lensing and abundance of {\redmapper} selected clusters by \cite{DESY1CL_2020_et_al} favored a cosmology that was in significant ($5.6\sigma$) tension with {\it{Planck}}. We presented a novel simulation-based framework to forward-model the impact of optical cluster selection on weak lensing data and applied it to dark matter simulations of a {\it{Planck}} cosmology. By forward modeling mock cluster selection, our model naturally predicts a scale-dependent lensing bias due to projection effects, while incorporating the impact of mass-dependent scatter in the mass--richness relation. Our findings are summarized as follows:
\begin{itemize}
\item Our simulation-based HOD model of cluster selection provides a statistically acceptable fit to the DES cluster abundance and weak lensing data in a {\it Planck} cosmology. This result resolves the tension found in the DES-Y1 analysis \cite{DESY1CL_2020_et_al}. 

\item Our best-fit model predicts significant richness-dependent enhancements to the cluster lensing profiles on large scales due to projection effects.  The predicted richness and scale dependence of this effect are in qualitative agreement with those measured using \redmapper\ clusters in the Buzzard simulation \cite{Wu_et_al_2022}.

\item Our best-fit model predicts a mass--richness relation with minimal redshift evolution.  The amplitude and slope are given by $\langle \log M_h | \lambda = 40, z = 0.35 \rangle = 14.25\pm0.027$ and  $d \log M_h / d \log \lambda = 1.52\pm0.132$. Notably, we find disagreement with the mass--richness relation of \cite{To_et_al_2021b}, despite the two works sharing part of their data vectors.

\end{itemize}

The most surprising aspect of our best-fit model is that it predicts a halo occupancy of only $\approx 3$ galaxies for halos of mass $10^{14}\ h^{-1}\ \Msun$, rising to 15 at $10^{14.5}\ h^{-1} \ \Msun$.  While cluster richness need not match halo occupancy, these are surprisingly low occupancy numbers. In our best-fit model, $\approx 80\%$ of the richness of a galaxy cluster is due to halos along the line of sight. While \cite{Myles_DES_et_al_2021} estimate a much lower projection fraction ($\approx 30\%$) in SDSS \redmapper\ clusters, a significant fraction of the galaxies that are spectroscopically associated with a cluster fall beyond the cluster's virial radius.  When we apply their method to our simulated catalogs we find projection fractions comparable to \cite{Myles_DES_et_al_2021}.  Nonetheless, the low occupancy of our best-fit model may be connected to the tension between our mass--richness relation and that of \cite{To_et_al_2021b}.  

Galaxy clusters have long been recognized as powerful cosmological probes of large-scale structure \cite[e.g.][]{Weinberg_PhR_2013}.  Optical clusters are, in principle, particularly powerful because of their relative abundance, but present formidable challenges to adequately characterizing their selection. Our paper demonstrates the feasibility of a simulation-based forward modeling strategy that could finally overcome this obstacle.  As such, our results are an important stepping stone on the way to realizing the promise of cluster cosmology in the era of Euclid, LSST, and the Roman Space Telescope.

\vspace{1em} \noindent \textbf{\emph{Acknowledgements.}} --- We thank Matteo Costanzi, Tim Eifler, and Elisabeth Krause for valuable discussions about this work. We thank the AbacusSummit team for making the simulation publicly available. Simulations were analyzed in part on computational resources of the Ohio Supercomputer Center \cite{OhioSupercomputerCenter1987}, with resources supported in part by the Center for Cosmology and AstroParticle Physics at the Ohio State University. We gratefully acknowledge the use of the {\sc{matplotlib}} software package \cite{Hunter_2007} and the GNU Scientific library \cite{GSL_2009}. This research has made use of the SAO/NASA Astrophysics Data System. ANS and ER received funding for this work from the Department of Energy (DOE) grant DE-SC0009913. ANS and TS received funding for this work from DOE grant DE-SC0020247 and the David and Lucile Packard Foundation. TS also received funding from the Japan Society for the Promotion of Science KAKENHI grant JP20H05855. HW and AL were supported by the DOE award DE-SC0021916 and the NASA award 15-WFIRST15-0008; AL was additionally supported by  Boise State's HERC Fellowship.  DHW acknowledges support of NSF grant AST-2009735. CHT received support from DOE grant DE-SC0011726.

\bibliographystyle{apsrev}
\bibliography{masterbib}

\newcommand{\noop}[1]{}
\begin{thebibliography}{46}
\expandafter\ifx\csname natexlab\endcsname\relax\def\natexlab#1{#1}\fi
\expandafter\ifx\csname bibnamefont\endcsname\relax
  \def\bibnamefont#1{#1}\fi
\expandafter\ifx\csname bibfnamefont\endcsname\relax
  \def\bibfnamefont#1{#1}\fi
\expandafter\ifx\csname citenamefont\endcsname\relax
  \def\citenamefont#1{#1}\fi
\expandafter\ifx\csname url\endcsname\relax
  \def\url#1{\texttt{#1}}\fi
\expandafter\ifx\csname urlprefix\endcsname\relax\def\urlprefix{URL }\fi
\providecommand{\bibinfo}[2]{#2}
\providecommand{\eprint}[2][]{\url{#2}}

\bibitem[{\citenamefont{{The Dark Energy Survey Collaboration}}(2005)}]{DES_2005}
\bibinfo{author}{\bibnamefont{{The Dark Energy Survey Collaboration}}}, \bibinfo{journal}{arXiv e-prints} \bibinfo{eid}{astro-ph/0510346} (\bibinfo{year}{2005}), \eprint{astro-ph/0510346}.

\bibitem[{\citenamefont{{Abbott} et~al.}(2020)\citenamefont{{Abbott}, {Aguena}, {Alarcon}, {Allam}, {Allen}, {Annis}, {Avila}, {Bacon}, {Bechtol}, {Bermeo} et~al.}}]{DESY1CL_2020_et_al}
\bibinfo{author}{\bibfnamefont{T.~M.~C.} \bibnamefont{{Abbott}}}, \bibinfo{author}{\bibfnamefont{M.}~\bibnamefont{{Aguena}}}, \bibinfo{author}{\bibfnamefont{A.}~\bibnamefont{{Alarcon}}}, \bibinfo{author}{\bibfnamefont{S.}~\bibnamefont{{Allam}}}, \bibinfo{author}{\bibfnamefont{S.}~\bibnamefont{{Allen}}}, \bibinfo{author}{\bibfnamefont{J.}~\bibnamefont{{Annis}}}, \bibinfo{author}{\bibfnamefont{S.}~\bibnamefont{{Avila}}}, \bibinfo{author}{\bibfnamefont{D.}~\bibnamefont{{Bacon}}}, \bibinfo{author}{\bibfnamefont{K.}~\bibnamefont{{Bechtol}}}, \bibinfo{author}{\bibfnamefont{A.}~\bibnamefont{{Bermeo}}}, \bibnamefont{et~al.}, \bibinfo{journal}{\prd} \textbf{\bibinfo{volume}{102}}, \bibinfo{eid}{023509} (\bibinfo{year}{2020}), \eprint{2002.11124}.

\bibitem[{\citenamefont{{Planck Collaboration} et~al.}(2016)\citenamefont{{Planck Collaboration}, {Ade}, {Aghanim}, {Arnaud}, {Ashdown}, {Aumont}, {Baccigalupi}, {Banday}, {Barreiro}, {Bartlett} et~al.}}]{Planck_2016}
\bibinfo{author}{\bibnamefont{{Planck Collaboration}}}, \bibinfo{author}{\bibfnamefont{P.~A.~R.} \bibnamefont{{Ade}}}, \bibinfo{author}{\bibfnamefont{N.}~\bibnamefont{{Aghanim}}}, \bibinfo{author}{\bibfnamefont{M.}~\bibnamefont{{Arnaud}}}, \bibinfo{author}{\bibfnamefont{M.}~\bibnamefont{{Ashdown}}}, \bibinfo{author}{\bibfnamefont{J.}~\bibnamefont{{Aumont}}}, \bibinfo{author}{\bibfnamefont{C.}~\bibnamefont{{Baccigalupi}}}, \bibinfo{author}{\bibfnamefont{A.~J.} \bibnamefont{{Banday}}}, \bibinfo{author}{\bibfnamefont{R.~B.} \bibnamefont{{Barreiro}}}, \bibinfo{author}{\bibfnamefont{J.~G.} \bibnamefont{{Bartlett}}}, \bibnamefont{et~al.}, \bibinfo{journal}{\aap} \textbf{\bibinfo{volume}{594}}, \bibinfo{eid}{A13} (\bibinfo{year}{2016}), \eprint{1502.01589}.

\bibitem[{\citenamefont{{Abbott} et~al.}(2018)\citenamefont{{Abbott}, {Abdalla}, {Alarcon}, {Allam}, {Annis}, {Avila}, {Aylor}, {Banerji}, {Banik}, {Baxter} et~al.}}]{DES_3x2pt_2018}
\bibinfo{author}{\bibfnamefont{T.~M.~C.} \bibnamefont{{Abbott}}}, \bibinfo{author}{\bibfnamefont{F.~B.} \bibnamefont{{Abdalla}}}, \bibinfo{author}{\bibfnamefont{A.}~\bibnamefont{{Alarcon}}}, \bibinfo{author}{\bibfnamefont{S.}~\bibnamefont{{Allam}}}, \bibinfo{author}{\bibfnamefont{J.}~\bibnamefont{{Annis}}}, \bibinfo{author}{\bibfnamefont{S.}~\bibnamefont{{Avila}}}, \bibinfo{author}{\bibfnamefont{K.}~\bibnamefont{{Aylor}}}, \bibinfo{author}{\bibfnamefont{M.}~\bibnamefont{{Banerji}}}, \bibinfo{author}{\bibfnamefont{N.}~\bibnamefont{{Banik}}}, \bibinfo{author}{\bibfnamefont{E.~J.} \bibnamefont{{Baxter}}}, \bibnamefont{et~al.}, \bibinfo{journal}{ArXiv e-prints}  (\bibinfo{year}{2018}), \eprint{1810.02322}.

\bibitem[{\citenamefont{{DES Collaboration} et~al.}(2021)\citenamefont{{DES Collaboration}, {Abbott}, {Aguena}, {Alarcon}, {Allam}, {Alves}, {Amon}, {Andrade-Oliveira}, {Annis}, {Avila} et~al.}}]{DES_3x2pt_2021}
\bibinfo{author}{\bibnamefont{{DES Collaboration}}}, \bibinfo{author}{\bibfnamefont{T.~M.~C.} \bibnamefont{{Abbott}}}, \bibinfo{author}{\bibfnamefont{M.}~\bibnamefont{{Aguena}}}, \bibinfo{author}{\bibfnamefont{A.}~\bibnamefont{{Alarcon}}}, \bibinfo{author}{\bibfnamefont{S.}~\bibnamefont{{Allam}}}, \bibinfo{author}{\bibfnamefont{O.}~\bibnamefont{{Alves}}}, \bibinfo{author}{\bibfnamefont{A.}~\bibnamefont{{Amon}}}, \bibinfo{author}{\bibfnamefont{F.}~\bibnamefont{{Andrade-Oliveira}}}, \bibinfo{author}{\bibfnamefont{J.}~\bibnamefont{{Annis}}}, \bibinfo{author}{\bibfnamefont{S.}~\bibnamefont{{Avila}}}, \bibnamefont{et~al.}, \bibinfo{journal}{arXiv e-prints} \bibinfo{eid}{arXiv:2105.13549} (\bibinfo{year}{2021}), \eprint{2105.13549}.

\bibitem[{\citenamefont{{Costanzi} et~al.}(2021)\citenamefont{{Costanzi}, {Saro}, {Bocquet}, {Abbott}, {Aguena}, {Allam}, {Amara}, {Annis}, {Avila}, {Bacon} et~al.}}]{Costanzi_et_al_2020}
\bibinfo{author}{\bibfnamefont{M.}~\bibnamefont{{Costanzi}}}, \bibinfo{author}{\bibfnamefont{A.}~\bibnamefont{{Saro}}}, \bibinfo{author}{\bibfnamefont{S.}~\bibnamefont{{Bocquet}}}, \bibinfo{author}{\bibfnamefont{T.~M.~C.} \bibnamefont{{Abbott}}}, \bibinfo{author}{\bibfnamefont{M.}~\bibnamefont{{Aguena}}}, \bibinfo{author}{\bibfnamefont{S.}~\bibnamefont{{Allam}}}, \bibinfo{author}{\bibfnamefont{A.}~\bibnamefont{{Amara}}}, \bibinfo{author}{\bibfnamefont{J.}~\bibnamefont{{Annis}}}, \bibinfo{author}{\bibfnamefont{S.}~\bibnamefont{{Avila}}}, \bibinfo{author}{\bibfnamefont{D.}~\bibnamefont{{Bacon}}}, \bibnamefont{et~al.}, \bibinfo{journal}{\prd} \textbf{\bibinfo{volume}{103}}, \bibinfo{eid}{043522} (\bibinfo{year}{2021}), \eprint{2010.13800}.

\bibitem[{\citenamefont{{To} et~al.}(2021{\natexlab{a}})\citenamefont{{To}, {Krause}, {Rozo}, {Wu}, {Gruen}, {DeRose}, {Rykoff}, {Wechsler}, {Becker}, {Costanzi} et~al.}}]{To_et_al_2021a}
\bibinfo{author}{\bibfnamefont{C.-H.} \bibnamefont{{To}}}, \bibinfo{author}{\bibfnamefont{E.}~\bibnamefont{{Krause}}}, \bibinfo{author}{\bibfnamefont{E.}~\bibnamefont{{Rozo}}}, \bibinfo{author}{\bibfnamefont{H.-Y.} \bibnamefont{{Wu}}}, \bibinfo{author}{\bibfnamefont{D.}~\bibnamefont{{Gruen}}}, \bibinfo{author}{\bibfnamefont{J.}~\bibnamefont{{DeRose}}}, \bibinfo{author}{\bibfnamefont{E.}~\bibnamefont{{Rykoff}}}, \bibinfo{author}{\bibfnamefont{R.~H.} \bibnamefont{{Wechsler}}}, \bibinfo{author}{\bibfnamefont{M.}~\bibnamefont{{Becker}}}, \bibinfo{author}{\bibfnamefont{M.}~\bibnamefont{{Costanzi}}}, \bibnamefont{et~al.}, \bibinfo{journal}{\mnras} \textbf{\bibinfo{volume}{502}}, \bibinfo{pages}{4093} (\bibinfo{year}{2021}{\natexlab{a}}), \eprint{2008.10757}.

\bibitem[{\citenamefont{{Wu} et~al.}(2022)\citenamefont{{Wu}, {Costanzi}, {To}, {Salcedo}, {Weinberg}, {Annis}, {Bocquet}, {Elidaiana da Silva Pereira}, {DeRose}, {Esteves} et~al.}}]{Wu_et_al_2022}
\bibinfo{author}{\bibfnamefont{H.-Y.} \bibnamefont{{Wu}}}, \bibinfo{author}{\bibfnamefont{M.}~\bibnamefont{{Costanzi}}}, \bibinfo{author}{\bibfnamefont{C.-H.} \bibnamefont{{To}}}, \bibinfo{author}{\bibfnamefont{A.~N.} \bibnamefont{{Salcedo}}}, \bibinfo{author}{\bibfnamefont{D.~H.} \bibnamefont{{Weinberg}}}, \bibinfo{author}{\bibfnamefont{J.}~\bibnamefont{{Annis}}}, \bibinfo{author}{\bibfnamefont{S.}~\bibnamefont{{Bocquet}}}, \bibinfo{author}{\bibfnamefont{M.}~\bibnamefont{{Elidaiana da Silva Pereira}}}, \bibinfo{author}{\bibfnamefont{J.}~\bibnamefont{{DeRose}}}, \bibinfo{author}{\bibfnamefont{J.}~\bibnamefont{{Esteves}}}, \bibnamefont{et~al.}, \bibinfo{journal}{arXiv e-prints} \bibinfo{eid}{arXiv:2203.05416} (\bibinfo{year}{2022}), \eprint{2203.05416}.

\bibitem[{\citenamefont{{Maksimova} et~al.}(2021)\citenamefont{{Maksimova}, {Garrison}, {Eisenstein}, {Hadzhiyska}, {Bose}, and {Satterthwaite}}}]{Maksimova_Summit_et_al_2021}
\bibinfo{author}{\bibfnamefont{N.~A.} \bibnamefont{{Maksimova}}}, \bibinfo{author}{\bibfnamefont{L.~H.} \bibnamefont{{Garrison}}}, \bibinfo{author}{\bibfnamefont{D.~J.} \bibnamefont{{Eisenstein}}}, \bibinfo{author}{\bibfnamefont{B.}~\bibnamefont{{Hadzhiyska}}}, \bibinfo{author}{\bibfnamefont{S.}~\bibnamefont{{Bose}}}, \bibnamefont{and} \bibinfo{author}{\bibfnamefont{T.~P.} \bibnamefont{{Satterthwaite}}}, \bibinfo{journal}{\mnras} \textbf{\bibinfo{volume}{508}}, \bibinfo{pages}{4017} (\bibinfo{year}{2021}), \eprint{2110.11398}.

\bibitem[{\citenamefont{{Rykoff} et~al.}(2014)\citenamefont{{Rykoff}, {Rozo}, {Busha}, {Cunha}, {Finoguenov}, {Evrard}, {Hao}, {Koester}, {Leauthaud}, {Nord} et~al.}}]{Rykoff_et_al_2014}
\bibinfo{author}{\bibfnamefont{E.~S.} \bibnamefont{{Rykoff}}}, \bibinfo{author}{\bibfnamefont{E.}~\bibnamefont{{Rozo}}}, \bibinfo{author}{\bibfnamefont{M.~T.} \bibnamefont{{Busha}}}, \bibinfo{author}{\bibfnamefont{C.~E.} \bibnamefont{{Cunha}}}, \bibinfo{author}{\bibfnamefont{A.}~\bibnamefont{{Finoguenov}}}, \bibinfo{author}{\bibfnamefont{A.}~\bibnamefont{{Evrard}}}, \bibinfo{author}{\bibfnamefont{J.}~\bibnamefont{{Hao}}}, \bibinfo{author}{\bibfnamefont{B.~P.} \bibnamefont{{Koester}}}, \bibinfo{author}{\bibfnamefont{A.}~\bibnamefont{{Leauthaud}}}, \bibinfo{author}{\bibfnamefont{B.}~\bibnamefont{{Nord}}}, \bibnamefont{et~al.}, \bibinfo{journal}{\apj} \textbf{\bibinfo{volume}{785}}, \bibinfo{eid}{104} (\bibinfo{year}{2014}), \eprint{1303.3562}.

\bibitem[{\citenamefont{{Drlica-Wagner} et~al.}(2018)\citenamefont{{Drlica-Wagner}, {Sevilla-Noarbe}, {Rykoff}, {Gruendl}, {Yanny}, {Tucker}, {Hoyle}, {Carnero Rosell}, {Bernstein}, {Bechtol} et~al.}}]{Drlica-Wagner_et_al_DESY1_2018}
\bibinfo{author}{\bibfnamefont{A.}~\bibnamefont{{Drlica-Wagner}}}, \bibinfo{author}{\bibfnamefont{I.}~\bibnamefont{{Sevilla-Noarbe}}}, \bibinfo{author}{\bibfnamefont{E.~S.} \bibnamefont{{Rykoff}}}, \bibinfo{author}{\bibfnamefont{R.~A.} \bibnamefont{{Gruendl}}}, \bibinfo{author}{\bibfnamefont{B.}~\bibnamefont{{Yanny}}}, \bibinfo{author}{\bibfnamefont{D.~L.} \bibnamefont{{Tucker}}}, \bibinfo{author}{\bibfnamefont{B.}~\bibnamefont{{Hoyle}}}, \bibinfo{author}{\bibfnamefont{A.}~\bibnamefont{{Carnero Rosell}}}, \bibinfo{author}{\bibfnamefont{G.~M.} \bibnamefont{{Bernstein}}}, \bibinfo{author}{\bibfnamefont{K.}~\bibnamefont{{Bechtol}}}, \bibnamefont{et~al.}, \bibinfo{journal}{\apjs} \textbf{\bibinfo{volume}{235}}, \bibinfo{eid}{33} (\bibinfo{year}{2018}), \eprint{1708.01531}.

\bibitem[{\citenamefont{{Zuntz} et~al.}(2018)\citenamefont{{Zuntz}, {Sheldon}, {Samuroff}, {Troxel}, {Jarvis}, {MacCrann}, {Gruen}, {Prat}, {S{\'a}nchez}, {Choi} et~al.}}]{Zuntz_et_al_DESY1_Metacal_2018}
\bibinfo{author}{\bibfnamefont{J.}~\bibnamefont{{Zuntz}}}, \bibinfo{author}{\bibfnamefont{E.}~\bibnamefont{{Sheldon}}}, \bibinfo{author}{\bibfnamefont{S.}~\bibnamefont{{Samuroff}}}, \bibinfo{author}{\bibfnamefont{M.~A.} \bibnamefont{{Troxel}}}, \bibinfo{author}{\bibfnamefont{M.}~\bibnamefont{{Jarvis}}}, \bibinfo{author}{\bibfnamefont{N.}~\bibnamefont{{MacCrann}}}, \bibinfo{author}{\bibfnamefont{D.}~\bibnamefont{{Gruen}}}, \bibinfo{author}{\bibfnamefont{J.}~\bibnamefont{{Prat}}}, \bibinfo{author}{\bibfnamefont{C.}~\bibnamefont{{S{\'a}nchez}}}, \bibinfo{author}{\bibfnamefont{A.}~\bibnamefont{{Choi}}}, \bibnamefont{et~al.}, \bibinfo{journal}{\mnras} \textbf{\bibinfo{volume}{481}}, \bibinfo{pages}{1149} (\bibinfo{year}{2018}), \eprint{1708.01533}.

\bibitem[{\citenamefont{{Hoyle} et~al.}(2018)\citenamefont{{Hoyle}, {Gruen}, {Bernstein}, {Rau}, {De Vicente}, {Hartley}, {Gaztanaga}, {DeRose}, {Troxel}, {Davis} et~al.}}]{Hoyle_et_al_DESY1_BPZ_2018}
\bibinfo{author}{\bibfnamefont{B.}~\bibnamefont{{Hoyle}}}, \bibinfo{author}{\bibfnamefont{D.}~\bibnamefont{{Gruen}}}, \bibinfo{author}{\bibfnamefont{G.~M.} \bibnamefont{{Bernstein}}}, \bibinfo{author}{\bibfnamefont{M.~M.} \bibnamefont{{Rau}}}, \bibinfo{author}{\bibfnamefont{J.}~\bibnamefont{{De Vicente}}}, \bibinfo{author}{\bibfnamefont{W.~G.} \bibnamefont{{Hartley}}}, \bibinfo{author}{\bibfnamefont{E.}~\bibnamefont{{Gaztanaga}}}, \bibinfo{author}{\bibfnamefont{J.}~\bibnamefont{{DeRose}}}, \bibinfo{author}{\bibfnamefont{M.~A.} \bibnamefont{{Troxel}}}, \bibinfo{author}{\bibfnamefont{C.}~\bibnamefont{{Davis}}}, \bibnamefont{et~al.}, \bibinfo{journal}{\mnras} \textbf{\bibinfo{volume}{478}}, \bibinfo{pages}{592} (\bibinfo{year}{2018}), \eprint{1708.01532}.

\bibitem[{\citenamefont{{Varga} et~al.}(2019)\citenamefont{{Varga}, {DeRose}, {Gruen}, {McClintock}, {Seitz}, {Rozo}, {Costanzi}, {Hoyle}, {MacCrann}, {Plazas} et~al.}}]{Varga_et_al_DESY1_2019}
\bibinfo{author}{\bibfnamefont{T.~N.} \bibnamefont{{Varga}}}, \bibinfo{author}{\bibfnamefont{J.}~\bibnamefont{{DeRose}}}, \bibinfo{author}{\bibfnamefont{D.}~\bibnamefont{{Gruen}}}, \bibinfo{author}{\bibfnamefont{T.}~\bibnamefont{{McClintock}}}, \bibinfo{author}{\bibfnamefont{S.}~\bibnamefont{{Seitz}}}, \bibinfo{author}{\bibfnamefont{E.}~\bibnamefont{{Rozo}}}, \bibinfo{author}{\bibfnamefont{M.}~\bibnamefont{{Costanzi}}}, \bibinfo{author}{\bibfnamefont{B.}~\bibnamefont{{Hoyle}}}, \bibinfo{author}{\bibfnamefont{N.}~\bibnamefont{{MacCrann}}}, \bibinfo{author}{\bibfnamefont{A.~A.} \bibnamefont{{Plazas}}}, \bibnamefont{et~al.}, \bibinfo{journal}{\mnras} \textbf{\bibinfo{volume}{489}}, \bibinfo{pages}{2511} (\bibinfo{year}{2019}), \eprint{1812.05116}.

\bibitem[{\citenamefont{{McClintock} et~al.}(2019)\citenamefont{{McClintock}, {Varga}, {Gruen}, {Rozo}, {Rykoff}, {Shin}, {Melchior}, {DeRose}, {Seitz}, {Dietrich} et~al.}}]{McClintock_DES_2019}
\bibinfo{author}{\bibfnamefont{T.}~\bibnamefont{{McClintock}}}, \bibinfo{author}{\bibfnamefont{T.~N.} \bibnamefont{{Varga}}}, \bibinfo{author}{\bibfnamefont{D.}~\bibnamefont{{Gruen}}}, \bibinfo{author}{\bibfnamefont{E.}~\bibnamefont{{Rozo}}}, \bibinfo{author}{\bibfnamefont{E.~S.} \bibnamefont{{Rykoff}}}, \bibinfo{author}{\bibfnamefont{T.}~\bibnamefont{{Shin}}}, \bibinfo{author}{\bibfnamefont{P.}~\bibnamefont{{Melchior}}}, \bibinfo{author}{\bibfnamefont{J.}~\bibnamefont{{DeRose}}}, \bibinfo{author}{\bibfnamefont{S.}~\bibnamefont{{Seitz}}}, \bibinfo{author}{\bibfnamefont{J.~P.} \bibnamefont{{Dietrich}}}, \bibnamefont{et~al.}, \bibinfo{journal}{\mnras} \textbf{\bibinfo{volume}{482}}, \bibinfo{pages}{1352} (\bibinfo{year}{2019}), \eprint{1805.00039}.

\bibitem[{\citenamefont{{Garrison} et~al.}(2018)\citenamefont{{Garrison}, {Eisenstein}, {Ferrer}, {Tinker}, {Pinto}, and {Weinberg}}}]{Garrison_et_al_2018}
\bibinfo{author}{\bibfnamefont{L.~H.} \bibnamefont{{Garrison}}}, \bibinfo{author}{\bibfnamefont{D.~J.} \bibnamefont{{Eisenstein}}}, \bibinfo{author}{\bibfnamefont{D.}~\bibnamefont{{Ferrer}}}, \bibinfo{author}{\bibfnamefont{J.~L.} \bibnamefont{{Tinker}}}, \bibinfo{author}{\bibfnamefont{P.~A.} \bibnamefont{{Pinto}}}, \bibnamefont{and} \bibinfo{author}{\bibfnamefont{D.~H.} \bibnamefont{{Weinberg}}}, \bibinfo{journal}{\apjs} \textbf{\bibinfo{volume}{236}}, \bibinfo{eid}{43} (\bibinfo{year}{2018}), \eprint{1712.05768}.

\bibitem[{\citenamefont{{Hadzhiyska} et~al.}(2022)\citenamefont{{Hadzhiyska}, {Eisenstein}, {Bose}, {Garrison}, and {Maksimova}}}]{Hadzhiyska_COMPASO_et_al_2022}
\bibinfo{author}{\bibfnamefont{B.}~\bibnamefont{{Hadzhiyska}}}, \bibinfo{author}{\bibfnamefont{D.}~\bibnamefont{{Eisenstein}}}, \bibinfo{author}{\bibfnamefont{S.}~\bibnamefont{{Bose}}}, \bibinfo{author}{\bibfnamefont{L.~H.} \bibnamefont{{Garrison}}}, \bibnamefont{and} \bibinfo{author}{\bibfnamefont{N.}~\bibnamefont{{Maksimova}}}, \bibinfo{journal}{\mnras} \textbf{\bibinfo{volume}{509}}, \bibinfo{pages}{501} (\bibinfo{year}{2022}), \eprint{2110.11408}.

\bibitem[{\citenamefont{{Tinker} et~al.}(2008)\citenamefont{{Tinker}, {Kravtsov}, {Klypin}, {Abazajian}, {Warren}, {Yepes}, {Gottl{\"o}ber}, and {Holz}}}]{Tinker_et_al_2008}
\bibinfo{author}{\bibfnamefont{J.}~\bibnamefont{{Tinker}}}, \bibinfo{author}{\bibfnamefont{A.~V.} \bibnamefont{{Kravtsov}}}, \bibinfo{author}{\bibfnamefont{A.}~\bibnamefont{{Klypin}}}, \bibinfo{author}{\bibfnamefont{K.}~\bibnamefont{{Abazajian}}}, \bibinfo{author}{\bibfnamefont{M.}~\bibnamefont{{Warren}}}, \bibinfo{author}{\bibfnamefont{G.}~\bibnamefont{{Yepes}}}, \bibinfo{author}{\bibfnamefont{S.}~\bibnamefont{{Gottl{\"o}ber}}}, \bibnamefont{and} \bibinfo{author}{\bibfnamefont{D.~E.} \bibnamefont{{Holz}}}, \bibinfo{journal}{\apj} \textbf{\bibinfo{volume}{688}}, \bibinfo{pages}{709} (\bibinfo{year}{2008}), \eprint{0803.2706}.

\bibitem[{\citenamefont{{Berlind} and {Weinberg}}(2002)}]{Berlind_2002}
\bibinfo{author}{\bibfnamefont{A.~A.} \bibnamefont{{Berlind}}} \bibnamefont{and} \bibinfo{author}{\bibfnamefont{D.~H.} \bibnamefont{{Weinberg}}}, \bibinfo{journal}{\apj} \textbf{\bibinfo{volume}{575}}, \bibinfo{pages}{587} (\bibinfo{year}{2002}), \eprint{astro-ph/0109001}.

\bibitem[{\citenamefont{{Zheng} et~al.}(2005)\citenamefont{{Zheng}, {Berlind}, {Weinberg}, {Benson}, {Baugh}, {Cole}, {Dav{\'e}}, {Frenk}, {Katz}, and {Lacey}}}]{Zheng_et_al_2005}
\bibinfo{author}{\bibfnamefont{Z.}~\bibnamefont{{Zheng}}}, \bibinfo{author}{\bibfnamefont{A.~A.} \bibnamefont{{Berlind}}}, \bibinfo{author}{\bibfnamefont{D.~H.} \bibnamefont{{Weinberg}}}, \bibinfo{author}{\bibfnamefont{A.~J.} \bibnamefont{{Benson}}}, \bibinfo{author}{\bibfnamefont{C.~M.} \bibnamefont{{Baugh}}}, \bibinfo{author}{\bibfnamefont{S.}~\bibnamefont{{Cole}}}, \bibinfo{author}{\bibfnamefont{R.}~\bibnamefont{{Dav{\'e}}}}, \bibinfo{author}{\bibfnamefont{C.~S.} \bibnamefont{{Frenk}}}, \bibinfo{author}{\bibfnamefont{N.}~\bibnamefont{{Katz}}}, \bibnamefont{and} \bibinfo{author}{\bibfnamefont{C.~G.} \bibnamefont{{Lacey}}}, \bibinfo{journal}{\apj} \textbf{\bibinfo{volume}{633}}, \bibinfo{pages}{791} (\bibinfo{year}{2005}), \eprint{astro-ph/0408564}.

\bibitem[{\citenamefont{{Navarro} et~al.}(1997)\citenamefont{{Navarro}, {Frenk}, and {White}}}]{NFW_1997}
\bibinfo{author}{\bibfnamefont{J.~F.} \bibnamefont{{Navarro}}}, \bibinfo{author}{\bibfnamefont{C.~S.} \bibnamefont{{Frenk}}}, \bibnamefont{and} \bibinfo{author}{\bibfnamefont{S.~D.~M.} \bibnamefont{{White}}}, \bibinfo{journal}{\apj} \textbf{\bibinfo{volume}{490}}, \bibinfo{pages}{493} (\bibinfo{year}{1997}), \eprint{astro-ph/9611107}.

\bibitem[{\citenamefont{{Correa} et~al.}(2015)\citenamefont{{Correa}, {Wyithe}, {Schaye}, and {Duffy}}}]{Correa_2015}
\bibinfo{author}{\bibfnamefont{C.~A.} \bibnamefont{{Correa}}}, \bibinfo{author}{\bibfnamefont{J.~S.~B.} \bibnamefont{{Wyithe}}}, \bibinfo{author}{\bibfnamefont{J.}~\bibnamefont{{Schaye}}}, \bibnamefont{and} \bibinfo{author}{\bibfnamefont{A.~R.} \bibnamefont{{Duffy}}}, \bibinfo{journal}{\mnras} \textbf{\bibinfo{volume}{452}}, \bibinfo{pages}{1217} (\bibinfo{year}{2015}), \eprint{1502.00391}.

\bibitem[{\citenamefont{{Angulo} et~al.}(2012)\citenamefont{{Angulo}, {Springel}, {White}, {Jenkins}, {Baugh}, and {Frenk}}}]{Angulo_et_al_2012}
\bibinfo{author}{\bibfnamefont{R.~E.} \bibnamefont{{Angulo}}}, \bibinfo{author}{\bibfnamefont{V.}~\bibnamefont{{Springel}}}, \bibinfo{author}{\bibfnamefont{S.~D.~M.} \bibnamefont{{White}}}, \bibinfo{author}{\bibfnamefont{A.}~\bibnamefont{{Jenkins}}}, \bibinfo{author}{\bibfnamefont{C.~M.} \bibnamefont{{Baugh}}}, \bibnamefont{and} \bibinfo{author}{\bibfnamefont{C.~S.} \bibnamefont{{Frenk}}}, \bibinfo{journal}{\mnras} \textbf{\bibinfo{volume}{426}}, \bibinfo{pages}{2046} (\bibinfo{year}{2012}), \eprint{1203.3216}.

\bibitem[{\citenamefont{{Busch} and {White}}(2017)}]{Busch_White_2017}
\bibinfo{author}{\bibfnamefont{P.}~\bibnamefont{{Busch}}} \bibnamefont{and} \bibinfo{author}{\bibfnamefont{S.~D.~M.} \bibnamefont{{White}}}, \bibinfo{journal}{\mnras} \textbf{\bibinfo{volume}{470}}, \bibinfo{pages}{4767} (\bibinfo{year}{2017}), \eprint{1702.01682}.

\bibitem[{\citenamefont{{Sunayama} et~al.}(2020)\citenamefont{{Sunayama}, {Park}, {Takada}, {Kobayashi}, {Nishimichi}, {Kurita}, {More}, {Oguri}, and {Osato}}}]{Sunayama_et_al_2020}
\bibinfo{author}{\bibfnamefont{T.}~\bibnamefont{{Sunayama}}}, \bibinfo{author}{\bibfnamefont{Y.}~\bibnamefont{{Park}}}, \bibinfo{author}{\bibfnamefont{M.}~\bibnamefont{{Takada}}}, \bibinfo{author}{\bibfnamefont{Y.}~\bibnamefont{{Kobayashi}}}, \bibinfo{author}{\bibfnamefont{T.}~\bibnamefont{{Nishimichi}}}, \bibinfo{author}{\bibfnamefont{T.}~\bibnamefont{{Kurita}}}, \bibinfo{author}{\bibfnamefont{S.}~\bibnamefont{{More}}}, \bibinfo{author}{\bibfnamefont{M.}~\bibnamefont{{Oguri}}}, \bibnamefont{and} \bibinfo{author}{\bibfnamefont{K.}~\bibnamefont{{Osato}}}, \bibinfo{journal}{\mnras} \textbf{\bibinfo{volume}{496}}, \bibinfo{pages}{4468} (\bibinfo{year}{2020}), \eprint{2002.03867}.

\bibitem[{\citenamefont{{Sunayama} and {More}}(2019)}]{Sunayama_More_2019}
\bibinfo{author}{\bibfnamefont{T.}~\bibnamefont{{Sunayama}}} \bibnamefont{and} \bibinfo{author}{\bibfnamefont{S.}~\bibnamefont{{More}}}, \bibinfo{journal}{\mnras} \textbf{\bibinfo{volume}{490}}, \bibinfo{pages}{4945} (\bibinfo{year}{2019}), \eprint{1905.07557}.

\bibitem[{\citenamefont{{Sunayama}}(2023)}]{Sunayama_et_al_2023}
\bibinfo{author}{\bibfnamefont{T.}~\bibnamefont{{Sunayama}}}, \bibinfo{journal}{\mnras} \textbf{\bibinfo{volume}{521}}, \bibinfo{pages}{5064} (\bibinfo{year}{2023}), \eprint{2205.03233}.

\bibitem[{\citenamefont{{Zeng} et~al.}(2023)\citenamefont{{Zeng}, {Salcedo}, {Wu}, and {Hirata}}}]{Zeng_et_al_2023}
\bibinfo{author}{\bibfnamefont{C.}~\bibnamefont{{Zeng}}}, \bibinfo{author}{\bibfnamefont{A.~N.} \bibnamefont{{Salcedo}}}, \bibinfo{author}{\bibfnamefont{H.-Y.} \bibnamefont{{Wu}}}, \bibnamefont{and} \bibinfo{author}{\bibfnamefont{C.~M.} \bibnamefont{{Hirata}}}, \bibinfo{journal}{\mnras} \textbf{\bibinfo{volume}{523}}, \bibinfo{pages}{4270} (\bibinfo{year}{2023}), \eprint{2210.16306}.

\bibitem[{\citenamefont{{Costanzi} et~al.}(2019)\citenamefont{{Costanzi}, {Rozo}, {Rykoff}, {Farahi}, {Jeltema}, {Evrard}, {Mantz}, {Gruen}, {Mandelbaum}, {DeRose} et~al.}}]{Costanzi_et_al_2019a}
\bibinfo{author}{\bibfnamefont{M.}~\bibnamefont{{Costanzi}}}, \bibinfo{author}{\bibfnamefont{E.}~\bibnamefont{{Rozo}}}, \bibinfo{author}{\bibfnamefont{E.~S.} \bibnamefont{{Rykoff}}}, \bibinfo{author}{\bibfnamefont{A.}~\bibnamefont{{Farahi}}}, \bibinfo{author}{\bibfnamefont{T.}~\bibnamefont{{Jeltema}}}, \bibinfo{author}{\bibfnamefont{A.~E.} \bibnamefont{{Evrard}}}, \bibinfo{author}{\bibfnamefont{A.}~\bibnamefont{{Mantz}}}, \bibinfo{author}{\bibfnamefont{D.}~\bibnamefont{{Gruen}}}, \bibinfo{author}{\bibfnamefont{R.}~\bibnamefont{{Mandelbaum}}}, \bibinfo{author}{\bibfnamefont{J.}~\bibnamefont{{DeRose}}}, \bibnamefont{et~al.}, \bibinfo{journal}{\mnras} \textbf{\bibinfo{volume}{482}}, \bibinfo{pages}{490} (\bibinfo{year}{2019}), \eprint{1807.07072}.

\bibitem[{\citenamefont{{Sinha} and {Garrison}}(2017)}]{Sinha_2017}
\bibinfo{author}{\bibfnamefont{M.}~\bibnamefont{{Sinha}}} \bibnamefont{and} \bibinfo{author}{\bibfnamefont{L.}~\bibnamefont{{Garrison}}}, \emph{\bibinfo{title}{{Corrfunc: Blazing fast correlation functions on the CPU}}}, \bibinfo{howpublished}{Astrophysics Source Code Library} (\bibinfo{year}{2017}), \eprint{1703.003}.

\bibitem[{\citenamefont{{Zhang} et~al.}(2019)\citenamefont{{Zhang}, {Jeltema}, {Hollowood}, {Everett}, {Rozo}, {Farahi}, {Bermeo}, {Bhargava}, {Giles}, {Romer} et~al.}}]{Zhang_DES_et_al_2019}
\bibinfo{author}{\bibfnamefont{Y.}~\bibnamefont{{Zhang}}}, \bibinfo{author}{\bibfnamefont{T.}~\bibnamefont{{Jeltema}}}, \bibinfo{author}{\bibfnamefont{D.~L.} \bibnamefont{{Hollowood}}}, \bibinfo{author}{\bibfnamefont{S.}~\bibnamefont{{Everett}}}, \bibinfo{author}{\bibfnamefont{E.}~\bibnamefont{{Rozo}}}, \bibinfo{author}{\bibfnamefont{A.}~\bibnamefont{{Farahi}}}, \bibinfo{author}{\bibfnamefont{A.}~\bibnamefont{{Bermeo}}}, \bibinfo{author}{\bibfnamefont{S.}~\bibnamefont{{Bhargava}}}, \bibinfo{author}{\bibfnamefont{P.}~\bibnamefont{{Giles}}}, \bibinfo{author}{\bibfnamefont{A.~K.} \bibnamefont{{Romer}}}, \bibnamefont{et~al.}, \bibinfo{journal}{Monthly Notices of the Royal Astronomical Society} \textbf{\bibinfo{volume}{487}}, \bibinfo{pages}{2578} (\bibinfo{year}{2019}), \eprint{1901.07119}.

\bibitem[{\citenamefont{{Wibking} et~al.}(2020)\citenamefont{{Wibking}, {Weinberg}, {Salcedo}, {Wu}, {Singh}, {Rodr{\'\i}guez-Torres}, {Garrison}, and {Eisenstein}}}]{Wibking_et_al_2020}
\bibinfo{author}{\bibfnamefont{B.~D.} \bibnamefont{{Wibking}}}, \bibinfo{author}{\bibfnamefont{D.~H.} \bibnamefont{{Weinberg}}}, \bibinfo{author}{\bibfnamefont{A.~N.} \bibnamefont{{Salcedo}}}, \bibinfo{author}{\bibfnamefont{H.-Y.} \bibnamefont{{Wu}}}, \bibinfo{author}{\bibfnamefont{S.}~\bibnamefont{{Singh}}}, \bibinfo{author}{\bibfnamefont{S.}~\bibnamefont{{Rodr{\'\i}guez-Torres}}}, \bibinfo{author}{\bibfnamefont{L.~H.} \bibnamefont{{Garrison}}}, \bibnamefont{and} \bibinfo{author}{\bibfnamefont{D.~J.} \bibnamefont{{Eisenstein}}}, \bibinfo{journal}{\mnras} \textbf{\bibinfo{volume}{492}}, \bibinfo{pages}{2872} (\bibinfo{year}{2020}), \eprint{1907.06293}.

\bibitem[{\citenamefont{{Salcedo} et~al.}(2022)\citenamefont{{Salcedo}, {Weinberg}, {Wu}, and {Wibking}}}]{Salcedo_et_al_2022}
\bibinfo{author}{\bibfnamefont{A.~N.} \bibnamefont{{Salcedo}}}, \bibinfo{author}{\bibfnamefont{D.~H.} \bibnamefont{{Weinberg}}}, \bibinfo{author}{\bibfnamefont{H.-Y.} \bibnamefont{{Wu}}}, \bibnamefont{and} \bibinfo{author}{\bibfnamefont{B.~D.} \bibnamefont{{Wibking}}}, \bibinfo{journal}{\mnras} \textbf{\bibinfo{volume}{510}}, \bibinfo{pages}{5376} (\bibinfo{year}{2022}), \eprint{2107.06314}.

\bibitem[{\citenamefont{{Foreman-Mackey} et~al.}(2013)\citenamefont{{Foreman-Mackey}, {Hogg}, {Lang}, and {Goodman}}}]{Foreman-Mackey_et_al_2016}
\bibinfo{author}{\bibfnamefont{D.}~\bibnamefont{{Foreman-Mackey}}}, \bibinfo{author}{\bibfnamefont{D.~W.} \bibnamefont{{Hogg}}}, \bibinfo{author}{\bibfnamefont{D.}~\bibnamefont{{Lang}}}, \bibnamefont{and} \bibinfo{author}{\bibfnamefont{J.}~\bibnamefont{{Goodman}}}, \bibinfo{journal}{\pasp} \textbf{\bibinfo{volume}{125}}, \bibinfo{pages}{306} (\bibinfo{year}{2013}), \eprint{1202.3665}.

\bibitem[{sup()}]{supp}
\bibinfo{howpublished}{See Supplemental Material at \url{URL_will_be_inserted_by_publisher} for full fiducial parameter posterior and comparison to literature richness-mass relations.}

\bibitem[{\citenamefont{{Grandis} et~al.}(2021)\citenamefont{{Grandis}, {Mohr}, {Costanzi}, {Saro}, {Bocquet}, {Klein}, {Aguena}, {Allam}, {Annis}, {Ansarinejad} et~al.}}]{Grandis_et_al_2021}
\bibinfo{author}{\bibfnamefont{S.}~\bibnamefont{{Grandis}}}, \bibinfo{author}{\bibfnamefont{J.~J.} \bibnamefont{{Mohr}}}, \bibinfo{author}{\bibfnamefont{M.}~\bibnamefont{{Costanzi}}}, \bibinfo{author}{\bibfnamefont{A.}~\bibnamefont{{Saro}}}, \bibinfo{author}{\bibfnamefont{S.}~\bibnamefont{{Bocquet}}}, \bibinfo{author}{\bibfnamefont{M.}~\bibnamefont{{Klein}}}, \bibinfo{author}{\bibfnamefont{M.}~\bibnamefont{{Aguena}}}, \bibinfo{author}{\bibfnamefont{S.}~\bibnamefont{{Allam}}}, \bibinfo{author}{\bibfnamefont{J.}~\bibnamefont{{Annis}}}, \bibinfo{author}{\bibfnamefont{B.}~\bibnamefont{{Ansarinejad}}}, \bibnamefont{et~al.}, \bibinfo{journal}{\mnras} \textbf{\bibinfo{volume}{504}}, \bibinfo{pages}{1253} (\bibinfo{year}{2021}), \eprint{2101.04984}.

\bibitem[{\citenamefont{{Baxter} et~al.}(2016)\citenamefont{{Baxter}, {Rozo}, {Jain}, {Rykoff}, and {Wechsler}}}]{Baxter_et_al_2016}
\bibinfo{author}{\bibfnamefont{E.~J.} \bibnamefont{{Baxter}}}, \bibinfo{author}{\bibfnamefont{E.}~\bibnamefont{{Rozo}}}, \bibinfo{author}{\bibfnamefont{B.}~\bibnamefont{{Jain}}}, \bibinfo{author}{\bibfnamefont{E.}~\bibnamefont{{Rykoff}}}, \bibnamefont{and} \bibinfo{author}{\bibfnamefont{R.~H.} \bibnamefont{{Wechsler}}}, \bibinfo{journal}{\mnras} \textbf{\bibinfo{volume}{463}}, \bibinfo{pages}{205} (\bibinfo{year}{2016}), \eprint{1604.00048}.

\bibitem[{\citenamefont{{Farahi} et~al.}(2016)\citenamefont{{Farahi}, {Evrard}, {Rozo}, {Rykoff}, and {Wechsler}}}]{Farahi_et_al_2016}
\bibinfo{author}{\bibfnamefont{A.}~\bibnamefont{{Farahi}}}, \bibinfo{author}{\bibfnamefont{A.~E.} \bibnamefont{{Evrard}}}, \bibinfo{author}{\bibfnamefont{E.}~\bibnamefont{{Rozo}}}, \bibinfo{author}{\bibfnamefont{E.~S.} \bibnamefont{{Rykoff}}}, \bibnamefont{and} \bibinfo{author}{\bibfnamefont{R.~H.} \bibnamefont{{Wechsler}}}, \bibinfo{journal}{\mnras} \textbf{\bibinfo{volume}{460}}, \bibinfo{pages}{3900} (\bibinfo{year}{2016}), \eprint{1601.05773}.

\bibitem[{\citenamefont{{Raghunathan} et~al.}(2019)\citenamefont{{Raghunathan}, {Patil}, {Baxter}, {Benson}, {Bleem}, {Chou}, {Crawford}, {Holder}, {McClintock}, {Reichardt} et~al.}}]{Raghunathan_et_al_2019}
\bibinfo{author}{\bibfnamefont{S.}~\bibnamefont{{Raghunathan}}}, \bibinfo{author}{\bibfnamefont{S.}~\bibnamefont{{Patil}}}, \bibinfo{author}{\bibfnamefont{E.}~\bibnamefont{{Baxter}}}, \bibinfo{author}{\bibfnamefont{B.~A.} \bibnamefont{{Benson}}}, \bibinfo{author}{\bibfnamefont{L.~E.} \bibnamefont{{Bleem}}}, \bibinfo{author}{\bibfnamefont{T.~L.} \bibnamefont{{Chou}}}, \bibinfo{author}{\bibfnamefont{T.~M.} \bibnamefont{{Crawford}}}, \bibinfo{author}{\bibfnamefont{G.~P.} \bibnamefont{{Holder}}}, \bibinfo{author}{\bibfnamefont{T.}~\bibnamefont{{McClintock}}}, \bibinfo{author}{\bibfnamefont{C.~L.} \bibnamefont{{Reichardt}}}, \bibnamefont{et~al.}, \bibinfo{journal}{\apj} \textbf{\bibinfo{volume}{872}}, \bibinfo{eid}{170} (\bibinfo{year}{2019}), \eprint{1810.10998}.

\bibitem[{\citenamefont{{Bleem} et~al.}(2020)\citenamefont{{Bleem}, {Bocquet}, {Stalder}, {Gladders}, {Ade}, {Allen}, {Anderson}, {Annis}, {Ashby}, {Austermann} et~al.}}]{Bleem_et_al_2020}
\bibinfo{author}{\bibfnamefont{L.~E.} \bibnamefont{{Bleem}}}, \bibinfo{author}{\bibfnamefont{S.}~\bibnamefont{{Bocquet}}}, \bibinfo{author}{\bibfnamefont{B.}~\bibnamefont{{Stalder}}}, \bibinfo{author}{\bibfnamefont{M.~D.} \bibnamefont{{Gladders}}}, \bibinfo{author}{\bibfnamefont{P.~A.~R.} \bibnamefont{{Ade}}}, \bibinfo{author}{\bibfnamefont{S.~W.} \bibnamefont{{Allen}}}, \bibinfo{author}{\bibfnamefont{A.~J.} \bibnamefont{{Anderson}}}, \bibinfo{author}{\bibfnamefont{J.}~\bibnamefont{{Annis}}}, \bibinfo{author}{\bibfnamefont{M.~L.~N.} \bibnamefont{{Ashby}}}, \bibinfo{author}{\bibfnamefont{J.~E.} \bibnamefont{{Austermann}}}, \bibnamefont{et~al.}, \bibinfo{journal}{\apjs} \textbf{\bibinfo{volume}{247}}, \bibinfo{eid}{25} (\bibinfo{year}{2020}), \eprint{1910.04121}.

\bibitem[{\citenamefont{{To} et~al.}(2021{\natexlab{b}})\citenamefont{{To}, {Krause}, {Rozo}, {Wu}, {Gruen}, {Wechsler}, {Eifler}, {Rykoff}, {Costanzi}, {Becker} et~al.}}]{To_et_al_2021b}
\bibinfo{author}{\bibfnamefont{C.}~\bibnamefont{{To}}}, \bibinfo{author}{\bibfnamefont{E.}~\bibnamefont{{Krause}}}, \bibinfo{author}{\bibfnamefont{E.}~\bibnamefont{{Rozo}}}, \bibinfo{author}{\bibfnamefont{H.}~\bibnamefont{{Wu}}}, \bibinfo{author}{\bibfnamefont{D.}~\bibnamefont{{Gruen}}}, \bibinfo{author}{\bibfnamefont{R.~H.} \bibnamefont{{Wechsler}}}, \bibinfo{author}{\bibfnamefont{T.~F.} \bibnamefont{{Eifler}}}, \bibinfo{author}{\bibfnamefont{E.~S.} \bibnamefont{{Rykoff}}}, \bibinfo{author}{\bibfnamefont{M.}~\bibnamefont{{Costanzi}}}, \bibinfo{author}{\bibfnamefont{M.~R.} \bibnamefont{{Becker}}}, \bibnamefont{et~al.}, \bibinfo{journal}{\prl} \textbf{\bibinfo{volume}{126}}, \bibinfo{eid}{141301} (\bibinfo{year}{2021}{\natexlab{b}}), \eprint{2010.01138}.

\bibitem[{\citenamefont{{Myles} et~al.}(2021)\citenamefont{{Myles}, {Alarcon}, {Amon}, {S{\'a}nchez}, {Everett}, {DeRose}, {McCullough}, {Gruen}, {Bernstein}, {Troxel} et~al.}}]{Myles_DES_et_al_2021}
\bibinfo{author}{\bibfnamefont{J.}~\bibnamefont{{Myles}}}, \bibinfo{author}{\bibfnamefont{A.}~\bibnamefont{{Alarcon}}}, \bibinfo{author}{\bibfnamefont{A.}~\bibnamefont{{Amon}}}, \bibinfo{author}{\bibfnamefont{C.}~\bibnamefont{{S{\'a}nchez}}}, \bibinfo{author}{\bibfnamefont{S.}~\bibnamefont{{Everett}}}, \bibinfo{author}{\bibfnamefont{J.}~\bibnamefont{{DeRose}}}, \bibinfo{author}{\bibfnamefont{J.}~\bibnamefont{{McCullough}}}, \bibinfo{author}{\bibfnamefont{D.}~\bibnamefont{{Gruen}}}, \bibinfo{author}{\bibfnamefont{G.~M.} \bibnamefont{{Bernstein}}}, \bibinfo{author}{\bibfnamefont{M.~A.} \bibnamefont{{Troxel}}}, \bibnamefont{et~al.}, \bibinfo{journal}{\mnras} \textbf{\bibinfo{volume}{505}}, \bibinfo{pages}{4249} (\bibinfo{year}{2021}), \eprint{2012.08566}.

\bibitem[{\citenamefont{{Weinberg} et~al.}(2013)\citenamefont{{Weinberg}, {Mortonson}, {Eisenstein}, {Hirata}, {Riess}, and {Rozo}}}]{Weinberg_PhR_2013}
\bibinfo{author}{\bibfnamefont{D.~H.} \bibnamefont{{Weinberg}}}, \bibinfo{author}{\bibfnamefont{M.~J.} \bibnamefont{{Mortonson}}}, \bibinfo{author}{\bibfnamefont{D.~J.} \bibnamefont{{Eisenstein}}}, \bibinfo{author}{\bibfnamefont{C.}~\bibnamefont{{Hirata}}}, \bibinfo{author}{\bibfnamefont{A.~G.} \bibnamefont{{Riess}}}, \bibnamefont{and} \bibinfo{author}{\bibfnamefont{E.}~\bibnamefont{{Rozo}}}, \bibinfo{journal}{\physrep} \textbf{\bibinfo{volume}{530}}, \bibinfo{pages}{87} (\bibinfo{year}{2013}), \eprint{1201.2434}.

\bibitem[{\citenamefont{{Ohio Supercomputer Center}}(1987)}]{OhioSupercomputerCenter1987}
\bibinfo{author}{\bibnamefont{{Ohio Supercomputer Center}}}, \emph{\bibinfo{title}{Ohio supercomputer center}}, \bibinfo{howpublished}{\url{http://osc.edu/ark:/19495/f5s1ph73}} (\bibinfo{year}{1987}).

\bibitem[{\citenamefont{Hunter}(2007)}]{Hunter_2007}
\bibinfo{author}{\bibfnamefont{J.~D.} \bibnamefont{Hunter}}, \bibinfo{journal}{Computing In Science \& Engineering} \textbf{\bibinfo{volume}{9}}, \bibinfo{pages}{90} (\bibinfo{year}{2007}).

\bibitem[{\citenamefont{Galassi et~al.}(2009)\citenamefont{Galassi, Davies, Theiler, Gough, Jungman, Alken, Booth, and Rossi}}]{GSL_2009}
\bibinfo{author}{\bibfnamefont{M.}~\bibnamefont{Galassi}}, \bibinfo{author}{\bibfnamefont{J.}~\bibnamefont{Davies}}, \bibinfo{author}{\bibfnamefont{J.}~\bibnamefont{Theiler}}, \bibinfo{author}{\bibfnamefont{B.}~\bibnamefont{Gough}}, \bibinfo{author}{\bibfnamefont{G.}~\bibnamefont{Jungman}}, \bibinfo{author}{\bibfnamefont{P.}~\bibnamefont{Alken}}, \bibinfo{author}{\bibfnamefont{M.}~\bibnamefont{Booth}}, \bibnamefont{and} \bibinfo{author}{\bibfnamefont{F.}~\bibnamefont{Rossi}}, \emph{\bibinfo{title}{GNU Scientific Library Reference Manual}}, \bibinfo{edition}{3rd} ed. (\bibinfo{year}{2009}).

\end{thebibliography}

\clearpage
\onecolumngrid
\appendix
\section{Supplemental Material}

\subsection{Full Fiducial Posterior}

\begin{figure}[ht]
\centering \includegraphics[width=0.50\textwidth]{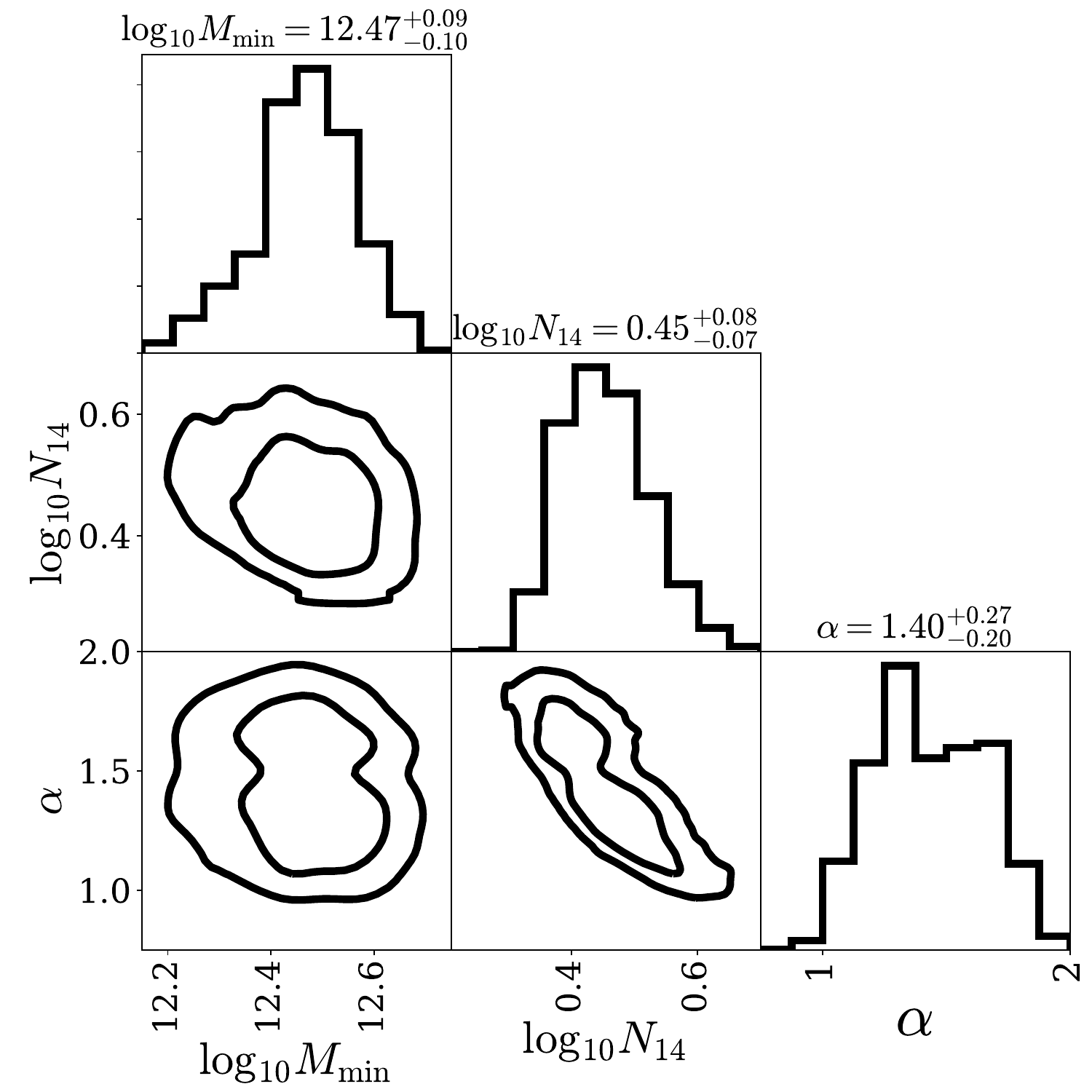}
    \caption{Posterior constraints on HOD and redshift evolution parameters from our MCMC analysis of the DES-Y1 cluster weak lensing data. Contours in each off-diagonal panel show the $68\%$ and $95\%$ confidence regions, and histograms in each diagonal panel show the 1D marginalized posterior distribution of each parameter, with values quoted above corresponding to the posterior mode and $68\%$ confidence intervals.} 
\label{fig:posterior}
\end{figure}

Figure \ref{fig:posterior} shows the full posterior distributions on our HOD parameters from our fiducial analysis of the DES-Y1 cluster weak lensing assuming {\it{Planck}} cosmology. 

\subsection{Comparison to Literature Richness-Mass Relations}

\begin{figure}[ht]
\centering \includegraphics[width=1.0\textwidth]{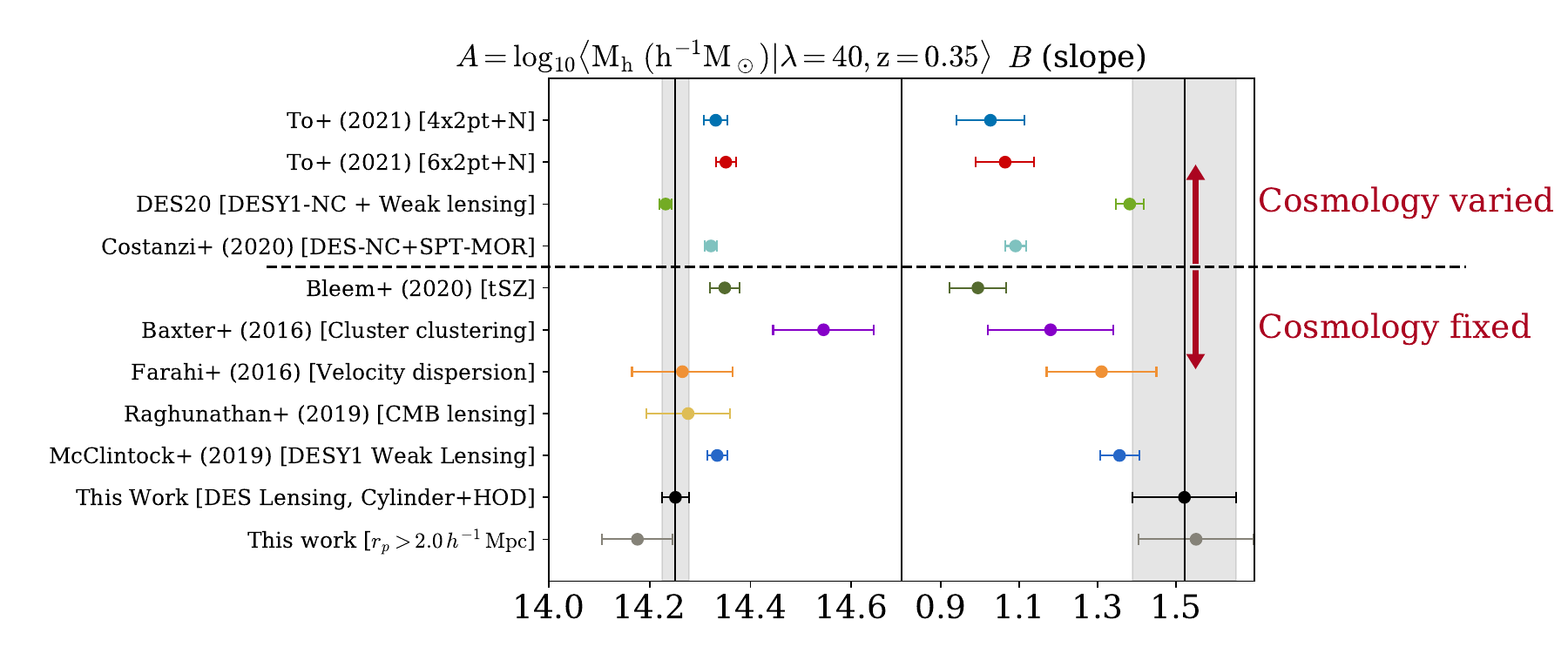}
    \caption{Comparison of the mean-mass of $\lambda = 40$ clusters at $z = 0.35$ (left-panel) and slope of the mass-richness relation (right-panel) from our work (black and grey points) with results from the literature \cite{Baxter_et_al_2016, Farahi_et_al_2016, Raghunathan_et_al_2019, Bleem_et_al_2020, DESY1CL_2020_et_al, Costanzi_et_al_2020, To_et_al_2021b}. All results are separated into those that vary cosmology (upper) and those that fix cosmology (lower).}
\label{fig:MOR}
\end{figure}

Figure \ref{fig:MOR} compares the inferred mass-richness relation from our fiducial analysis (black) and from our analysis excluding small scales ($r_p > 2.0 \, h^{-1} \, \mathrm{Mpc}$, grey) with results from the literature \cite{Baxter_et_al_2016, Farahi_et_al_2016, Raghunathan_et_al_2019, Bleem_et_al_2020, DESY1CL_2020_et_al, Costanzi_et_al_2020, To_et_al_2021b}. We parametrize the mass-richness relation as,
\beq
\langle M_h | \lambda \rangle = 10^A \times \left( \frac{\lambda}{40} \right)^{B} \, h^{-1} \, \mathrm{Mpc}
\eeq
where $A = \log_{10} \langle M_h | \lambda = 40, z = 0.35 \rangle$, the logarithmic mean mass of $\lambda = 40$ clusters at $z = 0.35$, and $B$ is the logarithmic slope in the mass-richness relation. Literature results are broken into those that vary cosmology (upper-group) and those that fix cosmology (lower-group). We see the aforementioned tension with the results of \cite{To_et_al_2021b} is not relaxed when we remove small scales from our data-vector. This implies that this tension is not driven by unmodeled small-scale physics such as baryon effects.

\end{document}